\PassOptionsToPackage{dvipsnames}{xcolor}
\documentclass{article} 
\usepackage{gmas_iclr2022_conference,times}

\usepackage{amsmath,amsfonts,bm}

\def\eqref#1{equation~\ref{#1}}

\def\1{\bm{1}}

\DeclareMathAlphabet{\mathsfit}{\encodingdefault}{\sfdefault}{m}{sl}
\SetMathAlphabet{\mathsfit}{bold}{\encodingdefault}{\sfdefault}{bx}{n}

\usepackage{hyperref}
\usepackage{url}

\usepackage[utf8]{inputenc} 
\usepackage[T1]{fontenc}    
\usepackage{hyperref}       
\usepackage{url}            
\usepackage{booktabs}       
\usepackage{amsfonts}       
\usepackage{nicefrac}       
\usepackage{microtype}      
\usepackage{amsmath}
\usepackage{algorithm}
\usepackage{algorithmicx}
\usepackage{algpseudocode}
\usepackage{graphicx}
\usepackage{subcaption}
\usepackage{amsmath}
\usepackage{amssymb}
\usepackage{textcomp}
\usepackage{xcolor}

\iclrfinalcopy
\title{Learning Truthful, Efficient, and Welfare Maximizing Auction Rules}

\author{Andrea Tacchetti, DJ Strouse, Marta Garnelo, Thore Graepel, \& Yoram Bachrach\\
DeepMind, UK\\
\texttt{\{atacchet, strouse, garnelo, thore, yorambac\}@deepmind.com}}

\begin{document}

\maketitle
\begin{abstract}

From social networks to supply chains, more and more aspects of how humans, firms and organizations interact is mediated by artificial learning agents. As the influence of machine learning systems grows, it is paramount that we study how to imbue our modern institutions with our own values and principles.
Here we consider the problem of allocating goods to buyers who have preferences over them in settings where the seller's aim is not to maximize their monetary gains, but rather to advance some notion of social welfare (e.g. the government trying to award construction licenses for hospitals or schools).
This problem has a long history in economics, and solutions take the form of auction rules. Researchers have proposed reliable auction rules that work in extremely general settings, and in the presence of information asymmetry and strategic buyers. However, these protocols require significant payments from participants resulting in low aggregate welfare. Here we address this shortcoming by casting auction rule design as a statistical learning problem, and trade generality for participant welfare effectively and automatically with a novel deep learning network architecture and auction representation. Our analysis shows that our auction rules outperform state-of-the art approaches in terms of participants welfare, applicability, robustness.
\end{abstract}


\section{Introduction}\label{sec:introduction}
More and more aspects of our lives are mediated by artificial learning agents; from social networks, to job hunting, and from route planning, to international trade, adaptive systems have become a centerpiece of modern institutions. As we manage the increased influence of artificial intelligence (AI), it is paramount that we are able to imbue our new institutions with our values, and trust them to implement detailed rules and protocols that embody these principles, even in complex, information-asymmetric scenarios with strategic participants.

Here we focus on the problem of designing a protocol for assigning bundles of goods or licenses to strategic buyers who have private valuations over them, and where the seller does not necessarily care about maximizing their proceeds from the sale, but is rather concerned with maximizing some notion of total participant welfare. 

This problem has a long history in economics, and solutions take the form of auctions protocols: after bidders report their valuation for the various bundles to the seller, the auction rule prescribes who gets which bundle, and how much each participant owes the seller. In particular, we highlight the Vickrey-Clarke-Groves (VCG) auction~\citep{clarke1971multipart,vickrey1961counterspeculation} which promotes truthful reports from buyers, and works in extremely general settings. VCG auctions, however, come at the cost of significant transfer from buyers to sellers, resulting in low aggregate participant welfare. This last observation has inspired VCG redistribution schemes, that is, modified VCG auction rules that recover some participant welfare trading away the general applicability of the original protocol~\citep{guo2010optimal}. These redistribution schemes, however, are hard to design, and often come with overly restrictive assumptions on participants preferences or behavior or on the nature of the goods up for sale.

Here we address these limitations by proposing a learning approach to auction rule design. We show that casting auction design as a learning problem allows us to trade setting generality for participant welfare effectively and automatically, and without need for overly restrictive assumptions (See Tab.~\ref{tab:comparison} for a qualitative comparison).

We start from the often reasonable assumption that bidders' valuations for the goods up for sale cannot take any value, but rather are sampled from an unknown and fixed probability distribution (e.g. it is very unlikely anyone would pay \$500,000 for a burrito). We introduce a representation of bidders' preferences and a network architecture that can be used to learn auction rules that a) incentivize truthful reports from participants, b) result in the social-welfare-maximizing allocation of the goods or bundles up for sale, and c) place minimal economic burden on participants (i.e. extract minimal payments).

We show that our approach can learn truthful mechanisms under a wide variety of settings, including various ``bidding languges''~\citep{nisan2000bidding} (i.e. the set or outcomes that bidders can have preferences over), arbitrary distributions of valuations, and arbitrary numbers of participants. Furthermore, our detailed analysis shows that the auction rules we learn outperform state-of-the art approaches to auction design in terms or participants welfare and robustness.

Auctions are a pillar of economics and remain the protocol of choice to allocate goods, services, and licenses in many applications world-wide. Here we show that, under reasonable assumptions, designing auctions that result in desirable allocations, and high participants' welfare can be cast as an optimization problem, and thus modern learning methods can be brought to bear. Our work provides an example of how we can imbue desirable values in learning agents and trust them to mediate complex interactions among humans, firms or other artificial agents in accordance to those values.

\section{Background and notation}\label{sec:bg-notation}
\begin{figure*}[ht]
    \centering
    \includegraphics[width=\textwidth]{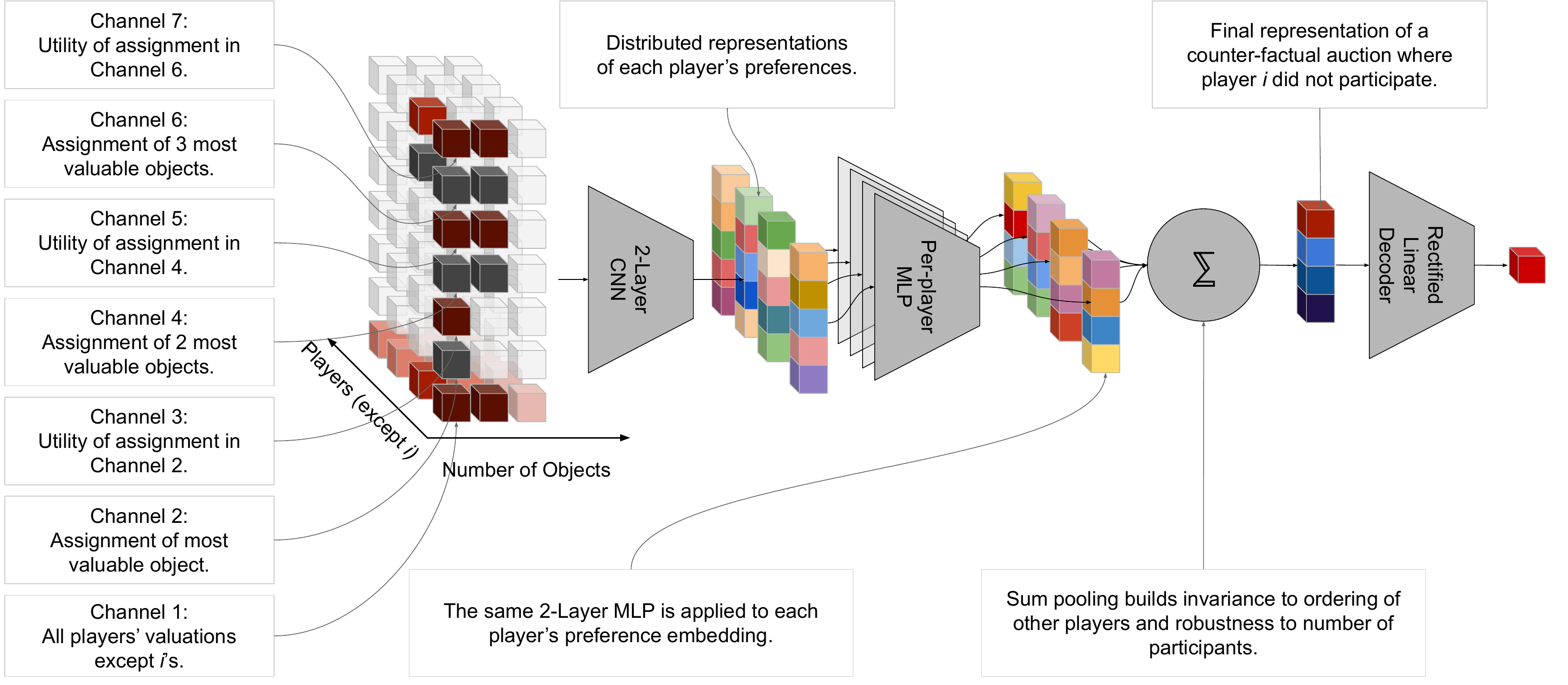}
    \caption{\small {\bf Example of Auction representation and network architecture} (best viewed in color). In this example we construct our auction representation for a multi-unit auction with decreasing marginal utilities with five participants, and three objects. The input tensor is of size $(n-1) \times |K|\times 2|K| + 1 = 4\times3\times7$. Darker shades of red indicate higher valuations. This representation is processed with a 2-layer CNN that extracts a per-player distributed representation of preferences and a 2-Layer MLP (shared weights across the players). The resulting embeddings are sum-pooled to build ordering invariance, and a rectified linear decoder outputs a single positive number as output.}\label{fig:auction-cnn}
\end{figure*}
In this section we provide some background, and introduce specific notation that will enable us to formally state the problem we tackle. 

{\bf Auction design}: Auctions are protocols to allocate bundles of goods to strategic buyers who have private valuations for them. The auction rules we consider prescribe that buyers report their preferences to the seller (e.g. telling them how much they value each bundle), at which point the seller uses an allocation rule $k$, and a payment rule $t$ to determine who gets which bundle, and how much each participant owes. Thus the auction rules we consider are fully specified by a choice of allocation rule and payment rule.

More formally, consider a set $I$ of items ordered absolutely (that is, everyone agrees on which is item 1, 2 and so on), and let $\mathcal{P}(I)$ be the power set of $I$ (i.e. the set of all possible bundles). An {\bf allocation} of the items is a function $k : N \rightarrow \mathcal{P}(I)$ mapping each participant $i$ to a bundle of items $k(i) \subseteq I$, such that for any $i \neq j$ we have $k(i) \cap k(j) = \emptyset$ (i.e. no item is allocated more than once), we further denote with $K$ the set of all possible allocations. A payment (or transfer) rule is a function $t: N \rightarrow \mathbb{R}$ that maps each player to an amount they owe the seller; payments can be negative indicating a transfer of money from the seller to the buyer. Finally, let $v_i: K \rightarrow \mathbb{R}_{+}$ be the valuation of participant $i$ for allocation $k$; as it is standard in the auction setting, we let $v_i$ only depend on the bundle assigned to $i$ (i.e. all allocations where $i$ gets items 1 and 3 have the same valuations: $i$ does not care who among participants $j$ and $w$ gets item 2).

{\bf Bidding languages}: Auction rules are often bespoke to specific settings, and the differentiating factor frequently resides in how bundles can be constructed and valued; bidding languages \citep{nisan2000bidding} allow us to formally specify these restrictions, and ensure that allocations remain computable. In this paper we consider 3 distinct bidding languages: 1) Multi-unit auctions with decreasing marginal utility: 2) Heterogeneous objects with unit demand, and 3) Hierarchical bundles.

In multi-unit auctions with decreasing marginal utilities, we assume that many indistinguishable units of the same product (e.g. oil barrels) are up for sale, and that participants valuation for a bundle only depends on how many units are in the bundle and not which ones, since all units are the same. Furthermore we assume that bundles with more units cannot be valued less than bundles with fewer units.

In auctions for heterogeneous objects with unit demand, we assume that various distinct products are available for sale (e.g. subscriptions to various cable channels), and that participants have distinct valuation for each individual item available. Furthermore, we let the valuation of a bundle coincide with that of its most valuable component (e.g. any bundle that includes HBO will be valued as much as HBO since participants can only watch one TV channel at the time).

Finally, in hierarchical bundles, we consider distinct products and allow participants to have valuations over specific groupings. Imagine that the items for sale are two pairs of trousers and two blazers, each matching one pair of trousers. We could let participants express their valuations for each item of clothing individually, for the two matching suits, or for all four items together, assuming no one is interested in purchasing mismatching suits. More formally, we arrange the items for sale on a binary tree and let participants express preferences for leaf nodes (i.e. individual objects), or any sub-tree.

{\bf Efficient allocation}: As we stated in the introduction, here we focus on constructing a protocol to allocate goods to strategic buyers that have preferences over them in pursuit of some notion of total welfare. It is thus a natural choice to allocate bundles ``efficiently'', that is, so as to maximize the total welfare of participants (before payments): $k^*=\arg \max_{k \in K} \sum_i v_i(k)$.

{\bf Strategic behavior}: Selecting the welfare maximizing allocation is difficult when the institutions we design do not have access to the true valuation profile of each participant, but rather can only trust what they report. This asymmetry in information leads to strategic behavior, that is, participants will report whatever preference $\theta_i$ maximizes their  utility $u_i$ under the auction rule (note that $u_i$ is a function of both allocation and payment). Formally, let $\theta_{-i}$ indicate all reports, truthful or otherwise, from all agents but $i$; then a strategic participant $i$ will report: $\theta_i = \arg\max_{\theta \in \Theta_i} u_i(k^*(\theta_{-i}, \theta), t(\theta_{-i}, \theta))$. In general $\theta_i \neq v_i$.

{\bf Truthful mechanisms}: In the presence of strategic participants, and for our choice of allocation function, it is possible to select a payment rule that makes reporting one's true preferences the dominant strategy. That is, for any agent $i$, and for all possible reports, or misreports, from other players $\theta_1,\ldots,\theta_{i-1},\theta_{i+1},\ldots,\theta_n$, the best course of action is to tell the truth: $\arg\max_{\theta \in \Theta_i} u_i(k^*(\theta_{-i}, \theta), t(\theta_{-i}, \theta)) = v_i$, $\forall \theta_{-i}$.

{\bf VCG mechanism and Groves payment rule}: We restrict our attention to auctions that are both efficient and truthful. All auction rules with these properties are  members of the Groves family, and their payment rule can be written as~\citep{groves1973incentives,green1979incentives,green1979social}:

\begin{equation}
\label{eq:groves-payment}
    t(i) = t(v_i, v_{-i}) = h( v_{-i} ) - \sum_{j \neq i} v_j( k^*(v_{-i}, v_i)),
\end{equation}
\noindent where, $h : \Theta_{-i} \rightarrow \mathbb{R}$ is any function that only depends on the reported types of agents other than $i$, and $k^*$ is the efficient allocation. Within this family, the VCG auction rule satisfies two further properties: 1) {\bf individual rationality}: buyers are never worse off by choosing to participate, i.e. $u_i \geq 0$, and 2) {\bf weak budget balance}: the seller does not need to subsidize the sale $\sum_i t(i) \geq 0$. The VCG auction is defined by the following choice of $h$: $h_{VCG}(\theta_{-i})=\sum_{j \neq i} v_j(k^*(\theta_{-i}))$ and resulting payment rule:
\begin{equation}
\label{eq:vcg-payment}
    t(v_i, v_{-i}) = \sum_{j \neq i} v_j(k^*(v_{-i})) - \sum_{j \neq i} v_j( k^*(v_{-i}, v_i)).
\end{equation}

{\bf Problem statement}\label{sec:problem-statement}
We aim to design {\it truthful}, and {\it efficient} auctions that {\it minimize the sum of payments collected by the seller}, while keeping the auction {\it individually-rational} and {\it weakly budget balanced}.

\section{Methods}\label{sec:methods}
\begin{table*}[ht]
    \centering
    \small
    \begin{tabular}{l | r r r r }
        \toprule
        \midrule
        Setting considered & \citeauthor{guo2010optimal} & \citeauthor{manisha2018learning} & G-CNN (ours) & R-CNN (ours)\\
        \toprule
        \midrule
	    No assumptions on $\rho$ & \textcolor{Bittersweet}{NO} & \textcolor{LimeGreen}{YES} & \textcolor{LimeGreen}{YES} & \textcolor{LimeGreen}{YES} \\
        $\rho$ is not known analytically  & \textcolor{Bittersweet}{NO} & \textcolor{LimeGreen}{YES} & \textcolor{LimeGreen}{YES} & \textcolor{LimeGreen}{YES}\\
        No restrictions on \# of participants & \textcolor{Bittersweet}{NO} & \textcolor{Bittersweet}{NO} & \textcolor{LimeGreen}{YES} & \textcolor{LimeGreen}{YES}\\
        Guarantees no-deficit & \textcolor{LimeGreen}{YES} & \textcolor{Bittersweet}{NO} & \textcolor{Bittersweet}{NO} & \textcolor{Bittersweet}{NO} \\
        Guarantees indiv. rationality & \textcolor{LimeGreen}{YES} & \textcolor{LimeGreen}{YES} & \textcolor{Bittersweet}{NO} & \textcolor{LimeGreen}{YES} \\        
        Multi-unit auctions & \textcolor{LimeGreen}{YES} & \textcolor{Bittersweet}{NO} & \textcolor{LimeGreen}{YES} & \textcolor{LimeGreen}{YES}\\
        Unit-demand auctions & \textcolor{Bittersweet}{NO} & \textcolor{LimeGreen}{YES} & \textcolor{LimeGreen}{YES} & \textcolor{LimeGreen}{YES}\\
        Hierarchical bundles auctions & \textcolor{Bittersweet}{NO} & \textcolor{Bittersweet}{NO} & \textcolor{LimeGreen}{YES} & \textcolor{LimeGreen}{YES}\\        
        \midrule
        \bottomrule
    \end{tabular}
    \caption{\small {\bf Qualitative results}. The method we present here can be applied in more general settings than previously proposed alternatives. Our models: G-CNN: learns a Groves payment rule directly using our data representation and network architecture. R-CNN: learns a VCG redistribution payment rule using our data representation and network architecture. $\rho$ indicates the distribution of valuation profiles.}\label{tab:comparison}
\end{table*}
We show how the problem of completing the Groves payment rule can be cast as a learning problem. We introduce our novel representation of efficient auctions, and a network architecture to learn minimum-payment, truthful auctions. We also point the reader to \citep{dutting2017optimal} for a related approach, and the literature around optimal auction design, which focuses on maximizing the seller's proceeds, rather than participants welfare~\citep{myerson1981optimal,riley1981optimal}.

\subsection{Loss function}
As stated in the introduction, we depart from the very general settings of the VCG auction by assuming that participants valuations are not arbitrary, but rather are sampled from a unknown, but fixed, probability distribution $\rho$.

Our objective is then equivalent to completing the payment rule $t(i)$ of a Groves mechanism so that, in expectation over valuation profiles sampled from $\rho$, we minimize the sum total of payments received by the mechanism. Minimizing payments without any further constraint will result in mechanisms that make $t(i)$ arbitrarily negative, and therefore require a subsidy to operate (i.e. they are not budget balanced). Thus we incorporate a {\it non-deficit} constraint. Similarly, we include an {\it individual rationality} constraint for all players. The resulting ``ideal'' mechanism design problem we wish to solve is:
\begin{align}
h^* = \arg\min_{h\in\mathcal{H}}\mathbb{E}_{v_i \sim \rho} \left[\sum_{i=1}^n t_i\right]\nonumber\\
\text{s.t. }\left[\sum_{i=1}^n t_i\right] \geq 0,\text{ and, } v_i(k^*) - t_i \geq 0,
\end{align}
where $t(i)$ is like in Eq.~\ref{eq:groves-payment}. As mentioned above, we assume we do not have access to the true distribution $\rho$, so that we cannot solve this minimization analytically. As is standard in statistical learning, we assume access to a data-set of $L$ $n$-player profiles $D = \{(v_1^l,\ldots,v_n^l | l = 1,\ldots,L\}$, sampled i.i.d. from $\rho$. We use Lagrange-like multipliers $\lambda_b$, and $\lambda_r$ to encode the non-deficit, and individual rationality constraints, and minimize the empirical version of our loss:

\begin{align}\label{eq:loss-function}
    \hat{h} = \arg\min_{h\in\mathcal{H}} \sum_{l=1}^L \sum_{i=1}^n t(i)^l\nonumber\\
    + \lambda_b\left(\min\left\{\sum_{i=1}^n t(i)^l, 0\right\}\right)^2\\
    +\lambda_r \sum_{i=1}^n\left(\left(\min\left\{v_i^l(k^*) - t(i)^l, 0\right\}\right)^2\right)\nonumber.
\end{align}

Concretely, we introduce two distinct ways to learn a Groves payments rule. The same representation, loss function and network architecture are used in both settings.

{\bf Selecting a Groves payment rule}: First, we investigate constructing a neural network to implement $\hat{h}$ directly and minimize the empirical loss in Eq.~\ref{eq:loss-function}, given a data-set of valuation profiles.

{\bf Learning a VCG redistribution mechanism}: Second, we learn a VCG {\it redistribution mechanism}. In this case, we use a neural network to implement a redistribution function $r(\cdot)$, and let $\hat{h}(\cdot) = h_{\text{VCG}}(\cdot) - r(\cdot)$\footnote{This is referred to as a ``redistribution'' mechanism since it can be viewed as collecting the VCG payments before ``redistributing'' some money back to participants.}. Note that in this case individual rationality can be guaranteed by simply ensuring that $r(\cdot)$ takes non-negative values, since VCG is already individually rational.

\subsection{Auction representation}\label{sec:hypothesis-space}

{\bf Representing auctions} We introduce a novel representation of auctions that supports learning Groves payment rules with Deep Neural Networks. Fig.~\ref{fig:auction-cnn} shows an example of our representation and architecture for an auction with three objects and five participants.

First we note that when computing $t(i)$ the function we wish to learn takes as input valuations from ``other'' players $v_{-i}$, and has no knowledge of player $i$'s profile (see Eq.~\ref{eq:groves-payment}). We construct our representation as follows: first, we construct an ``allocation oracle'' to compute the efficient allocation $k^*$ for any set of valuations (this is easy to construct given our choice of bidding languages; see~\citep{nisan2000bidding} for details on how to construct such an oracle). Second, we choose to represent each of the $v_{-i}$ as outcomes of $|K|$ counter-factual auctions, each for the most valuable $p$ bundles with $p=1,\ldots,|K|$. The idea here is to provide information about the relative rank of each bundle valuation.

Precisely, given a data-set of realized valuation profiles $D$, and an allocation oracle, we construct, for each player $i$ a tensor with shape $|K|\times (n-1)\times 2|K| + 1$. Each $|K|\times (n-1)$ slice contains matrix $V_{-i} \in \mathbb{R}_{+}^{|K| \times (n-1)}$ with non-negative entries $(m,j)$ representing the valuation of player $j$ for bundle $m$ (that is $v_i(k_m)$, where $k_m$ allocates $m$, and nothing else, to $j$). Each successive channel $p$ is constructed by considering a counter-factual auction where the $n-1$ players bid for the $p$ most valuable bundles. In particular, the second channel contains the allocation matrix $k_1^* \in \{0,1\}^{|K| \times (n-1)}$ with entries $(m,j) = 1$ if bidder $j$ is allocated bundle $m$ in this auction, and zero otherwise. The third channel represents the amount of utility realized by each player for this allocation before payments (i.e. the element-wise product between the first and second channels). Similarly, the fourth channel contains $k_2^*$: the allocation for two bundles, and the fifth channel contains the element-wise product between channels 1 and 4, and so on until all bundles are considered. We alter this representation slightly to in multi-unit auctions with decreasing marginal utilities. In this case we let $V_{-i}$ be a matrix with shape $|B| \times (n-1)$, with $B$ the set of available items, and containing, for each player, the marginal utility of adding one item to their bundle.

{\bf A network architecture to learn Groves payment rules} Given our auction representation, we propose an architecture to learn a Groves payment rule that satisfies the following: a) {\it anonymity}: the same exact function is applied to each player, b) robustness to ordering: $t(i)$ does not change if players $j$ and $w$ swap valuations.

For each player $i$, we construct the input tensor of size $|K| \times (n-1) \times (2|K| + 1)$ described above and pass it through a $2$-layer CNN. The first layer uses $64$ filters of spatial size $1\times1$ so as to construct an embedding of each individual bid (how soon each bundle is allocated, and how much utility it realizes can be readily extracted from a single ``column'' in our representation). The second CNN layer has $64$ filters of size $|K|\times 1$. The CNN's output thus has size $1 \times (n - 1) \times 64$, and contains an embedding of each of the $n-1$ players' preferences. We follow our CNN with a 2-Layer, $64$ hidden and output units MLP, which we apply independently to each of the $(n-1)$ player preference embeddings to produce a new embedding for each player. We then sum-pool over the $n-1$ players (which guarantees the desired robustness properties), and apply a linear decoder (with ReLU rectification) to output a single value for either $\hat{h}$ directly, or for a redistribution function $r$. It is worth noting that this architecture is a DeepSets network applied to a graph of $n-1$ nodes with a single global output, where node functions are our CNN+MLP and the aggregator function is a sum~\citep{manzil2017deepsets,battaglia2018relational}.

\subsection{Experimental procedure and baselines}
{\bf Baselines} We consider four baselines. 1) {\it VCG auctions}, the most commonly used Groves mechanism: a truthful, efficient, weakly budget balanced and individually rational auction. 2) {\it\cite{guo2010optimal}} a provably optimal-in-expectation linear VCG redistribution mechanism, which requires $n<|K|$, analytical knowledge of $\rho$, and only handles multi-unit auctions. 3) {\it\cite{manisha2018learning}} a VCG redistribution learned using a MLP architecture that requires $n<|K|$, and only works with unit-demand valuations. 4) {\it MLP based architecture} lastly, we compare to a 2-layer, 128-hidden-unit MLP that operates on a flattened version of the same data as our method to empirically support our choice of representation and architecture.

{\bf Experimental procedure} For each combination of number of participants, valuation distribution and bidding language considered, we construct sample auctions (i.e. valuation profiles for all participants, expressed in the appropriate language) and collect training and validation data-sets containing $100{,}000$ and $2{,}000$ auctions respectively. For each auction, we construct the representation described in Sec.~\ref{sec:hypothesis-space}, and train the auction design network above using Adam SGD \citep{kingma2015adam} with a learning rate of $10^{-5}$, mini-batches of size $256$, and for $250{,}000$ iterations. In all experiments we set $\lambda_b = \lambda_r = 100$ (see Eq.~\ref{eq:loss-function}). After training, we use our held-out test set to report performance. The number of objects for sale were as follows: with non-decreasing marginal utilities: 15 objects, with heterogeneous objects and unit-demand: 8 objects, and with hierarchical bundles: 8 component objects (resulting in 15 bundles).

\section{Results}
\begin{figure*}[ht]
    \centering
    \begin{subfigure}[b]{\textwidth}
    \includegraphics[width=\textwidth]{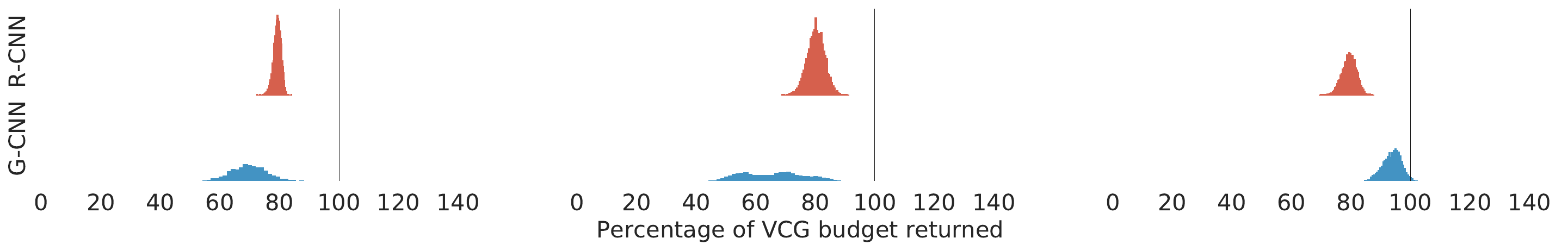}
    \caption{\small $\rho = \mathcal{N}(\mathcal{N}(10.0, 1.0), \mathcal{N}(2.0, 0.5))$. Train: $n\in\{9,11\}$. Test: $n=10$. {\bf Left}: multi-unit auction with decreasing marginal utilities. {\bf Middle}: Heterogeneous objects with unit demand. {\bf Right}: hierarchical bundles.}\label{fig:robust-to-n}
    \end{subfigure}
    
    \begin{subfigure}[b]{\textwidth}
    \includegraphics[width=\textwidth]{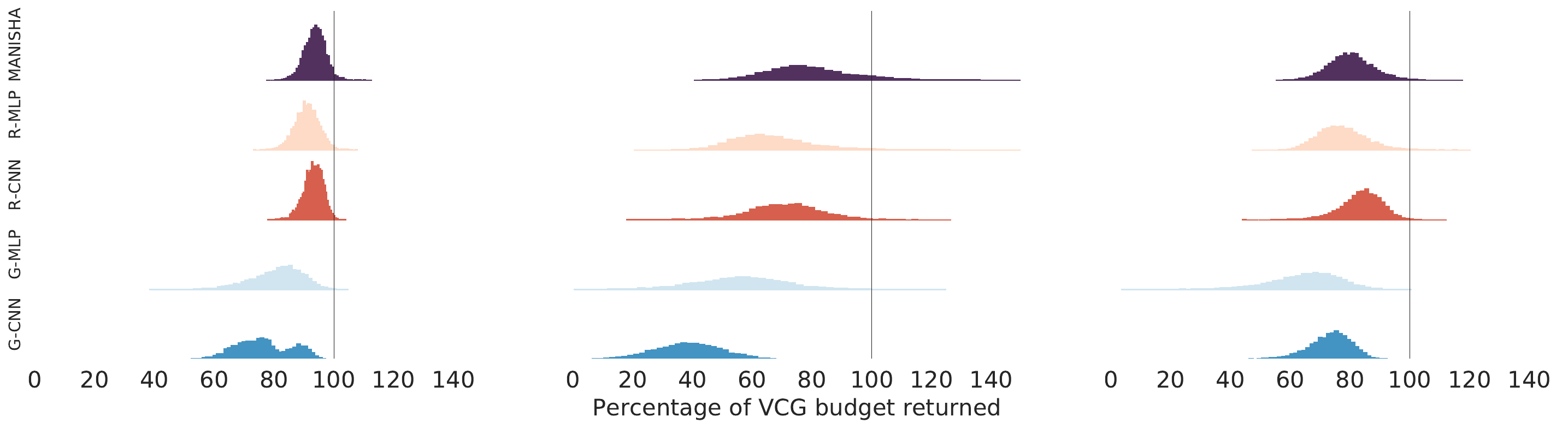}
    \caption{\small{\bf Left}: $\rho = \mathcal{N}(\mathcal{N}(10.0, 1.0), \mathcal{N}(2.0, 0.5))$. {\bf Middle}: $\rho=\mathcal{N}(10.0, 2.0)$, {\bf Right}: $\rho = \mathcal{U}(0.0, 1.0)$. $n=10$. Bidding language: Heterogeneous objects with unit-demand.}\label{fig:manisha}    
    \end{subfigure}
    \caption{\small Each plot shows a normalized count of how many auctions (among the $2000$ we used for testing) resulted in the fraction of the VCG payments reported on the horizontal axis being ``returned'' to the participants. A VCG auction would result in a score of $0\%$, since payments are collected and nothing is redistributed. A score of $100\%$ indicates that the auction is ``perfectly budged balanced'' meaning no net payments are extracted. The goal of our design is to construct payment rules that concentrate the redistribution as close as possible to $100\%$ without ever exceeding it: concentrations around high redistribution scores indicate that participant achieve high aggregate welfare (i.e. low net payments), while redistributions exceeding $100\%$ indicate that the auctioneer incurred a deficit (i.e. the weak budget balance constrained is violated). In both figures R-CNN refers to learning a VCG redistribution scheme with our representation and architecture, whereas G-CNN refers to the case where we learn a payment rule directly (see Sec.~\ref{sec:bg-notation}). Fig.~\ref{fig:robust-to-n} shows results for a specific choice of $\rho$, and three bidding languages. Fig~\ref{fig:manisha} includes results for three choices of $\rho$, and a fixed bidding language, and compares the outcome of our auctions to alternative designs: MANISHA refers to the method outlined in~\protect\cite{manisha2018learning}, and R-MLP and G-MLP were constructed by using the same exact loss functions and data as R-CNN and G-CNN respectively, but using a flat representation and an MLP network (see Sec.~\ref{sec:methods}).}\label{fig:quant-results}
\end{figure*}

{\bf Qualitative comparison with alternative methods} We start with a qualitative comparison with two existing alternative methods to automatically construct VCG redistribution protocols, and highlight how our method can be applied in more general settings in Tab.~\ref{tab:comparison}. A quantitative comparison with these two methods (in the settings in which they can be applied) shows how our methods also leads to better performance in practice. Importantly, while our method does not guarantee we will find auctions that are weakly budget balanced and individually rational, our quantitative results show that, in practice, we find zero, or next-to-zero violations of these constraints, and in particular the individual rationality constraint is never violated (this is expected since we are minimizing payments, thus making participation more appealing).

{\bf Quantitative results} We illustrate quantitative results on synthetic auction data-sets in Fig.~\ref{fig:quant-results}.

Fig.~\ref{fig:robust-to-n} shows the performance of auctions learned using our two methods G-CNN (where we learn a Groves payment rule directly), and R-CNN (where we learn a redistribution mechanism). It is clear how the auctions we learn result in high redistributions with minimal budget balance violation (none for R-CNN). For this experiment, valuations were sampled from a Gaussian with mean and standard deviations sampled from two independent Gaussian distrbutions $\rho = \mathcal{N}(\mathcal{N}(10.0, 1.0), \mathcal{N}(2.0, 0.5))$. Furthermore, this experiment showcases the high applicability of our method (the only method to support all three bidding languages), and our architecture's ability to interpolate to unseen number of participants: training was performed using auctions with either $9$ or $11$ participants, while testing used auctions with $10$ participants; no other method supports this transfer learning.

Fig.~\ref{fig:manisha} shows a comparison between our method, an MLP network that operates on the same data, so as to validate our choice of data representation and network architecture, and the work from~\cite{manisha2018learning}. The distributions of valuations we used for this experiment where a Gaussian with mean and standard deviations sampled from two independent Gaussian distributions $\rho = \mathcal{N}(\mathcal{N}(10.0, 1.0), \mathcal{N}(2.0, 0.5))$, a Gaussian with fixed mean and standard deviation: $\rho=\mathcal{N}(10.0, 2.0)$, and the uniform distribution: $\rho = \mathcal{U}(0.0, 1.0)$. Only unit demand auctions with $n=10$ participants were used since the work from Manisha et al. requires $n<|K|$ and does not support other bidding languages.

We end this section with a quantitative comparison with the provably optimal in expectation redistribution scheme of~\cite{guo2010optimal}, in the only setting where it is applicable: multi-unit auctions with decreasing marginal utility with valuation sampled from the uniform distribution, and $n<|K|$.  Specifically we consider two settings, and report performance as fractions of the aggregate VCG payments returned (we report redistribution mean and standard deviation exclusively since the baseline results are pulled directly from the original paper). With $n=3$ R-CNN redistributes $80\pm1\%$ of VCG payments with a deficit of $0\pm0\%$, G-CNN redistributes $87\pm1\%$ of VCG payments with a deficit of $0\pm1\%$, while the redistribiton scheme of~\cite{guo2010optimal} redistributes $76\%$ of VCG payments, and guarantees no deficit. With $n=7$ R-CNN redistributes $84\pm6$ of VCG payments with a deficit of $0\pm0\%$, G-CNN redistributes $95\pm0\%$ of VCG payments with a deficit of $0\pm1\%$, while the redistribiton scheme of~\cite{guo2010optimal} redistributes $94\%$ of VCG payments and guarantees no deficit (mean and std. dev. over $2000$ held out synthetic auctions for G-CNN and R-CNN).

Our experiments show that auctions learned using our statistical learning formulation, data representation and network architecture result in a significantly smaller economic burden on the participants than alternative designs, and crucially, that we are able to learn auction rules with a better trade-off between participant welfare and budget balance violations. Furthermore, our experiments showcase the wide applicability and robustness of our method with respect to choice of bidding languages, number of participants, distribution of valuations, and interpolation to unseen scenarios.

\section{Discussion}\label{sec:discussion}

We introduced a novel way to represent auctions, and proposed a neural architecture, to learn truthful and efficient auctions with minimal economic burden on the participants. Our methods can be applied on a wide variety of settings including arbitrary distributions, complex bidding languages and variable number of participants. Our empirical analysis shows how the resulting auctions yields high participants' welfare and almost never require a subsidy. Moreover, restricting our auction designs to the Groves family provides a template for constructing adaptive systems that remain firmly planted in the theoretical foundations of economics and mechanism design.

Auction design is a pillar of economics and social sciences and the domain of choice to study how a institution can mediate the interactions of strategic participants in pursuit of group-level aspirations (e.g. maximize aggregate welfare). Nonetheless, very few attempts to apply machine learning ideas to this setting have been made. Here we have shown that, under reasonable assumptions, auction design can be turned into a statistical learning problem and modern methods can be brought to bear.
The recent renaissance of Artificial Intelligence points to a future where institutions are largely built around adaptive systems, and where we must entrust learning agents with the automatic translation of high-level directives into low-level incentive structures and interaction rules.


\newpage

\bibliographystyle{gmas_iclr2022_conference}
\bibliography{references}

\end{document}